# Relaxation-based viscosity mapping for magnetic particle imaging

M Utkur[1,2], Y Muslu[1,2] and E U Saritas[1,2,3]

[1] Department of Electrical and Electronics Engineering, Bilkent University, Ankara, Turkey

[2] National Magnetic Resonance Research Center (UMRAM), Bilkent University, Ankara, Turkey

[3] Neuroscience Graduate Program, Bilkent University, Ankara, Turkey

E-mail: mustafa.utkur@bilkent.edu.tr

**Abstract**

Magnetic Particle Imaging (MPI) has been shown to provide remarkable contrast for imaging applications such as angiography, stem cell tracking, and cancer imaging. Recently, there is growing interest in the functional imaging capabilities of MPI, where "color MPI" techniques have explored separating different nanoparticles, which could potentially be used to distinguish nanoparticles in different states or environments. Viscosity mapping is a promising functional imaging application for MPI, as increased viscosity levels in vivo have been associated with numerous diseases such as hypertension, atherosclerosis, and cancer. In this work, we propose a viscosity mapping technique for MPI through the estimation of the relaxation time constant of the nanoparticles. Importantly, the proposed time constant estimation scheme does not require any prior information regarding the nanoparticles. We validate this method with extensive experiments in an in-house magnetic particle spectroscopy (MPS) setup at four different frequencies (between 250 Hz and 10.8 kHz) and at three different field strengths (between 5 mT and 15 mT) for viscosities ranging between 0.89 mPa•s to 15.33 mPa•s. Our results demonstrate the viscosity mapping ability of MPI in the biologically relevant viscosity range.

## 1. Introduction

Magnetic Particle Imaging (MPI) is a rapidly developing imaging modality (Gleich and Weizenecker 2005, Goodwill *et al* 2012, Saritas *et al* 2013a, Bauer *et al* 2015) with potential biomedical applications such as angiography (Weizenecker *et al* 2009, Lu *et al* 2013), stem cell tracking (Zheng *et al* 2015, 2016, Them *et al* 2016), cancer imaging (Yu *et al* 2016), guiding cardiovascular interventions (Haegele *et al* 2012), and predicting the effectiveness of magnetic hyperthermia (Murase *et al* 2015, Hensley *et al* 2016). MPI detects the response of superparamagnetic iron oxide (SPIO) nanoparticles to applied external magnetic fields, without any signal from the background tissue. This response is sensitive to the local environment of the SPIOs such as the viscosity of the medium that they are in or their binding state to chemicals, showing a remarkable potential for functional imaging with MPI.

One potential functional imaging application for MPI is in vivo viscosity mapping. In literature, blood viscosity was shown to change with hematocrit level (Pirofsky 1953), and several studies associated high hematocrit levels in the blood plasma with important diseases such as hypertension (Tibblin *et al* 1966), cerebral infarction (Tohgi *et al* 1978), angina pectoris (Burch and Depasquale 1965), ischemic heart disease (Fuchs *et al* 1990, 1984), and extensive coronary artery disease (Lowe *et al* 1980). High blood viscosity levels were also found to be prognostic of certain cancer types (von Tempelhoff *et al* 2002, Chandler and Schmer 1986). It was further suggested that cancer cells have increased cellular viscosity when compared to healthy cells (Guyer and Claus 1942). Importantly, for cell tracking applications, viscosity in a cell cytoplasm can give information regarding the uptake of external particles into the cell environment as well as activities occurring within the cell. Interactions in the cell environment such as signaling, transportation of molecules, enzymatic activity, binding effect were interrelated with viscosity (Williams *et al* 1997, Rauwerdink and Weaver 2010a, 2011, Giustini *et al* 2012). Diagnoses of atherosclerosis (Deliconstantinos *et al* 1995), diabetes (Nadiv *et al* 1994), Alzheimer's disease (Zubenko *et al* 1999), and neurological disorders (Fahey *et al* 1965) were found to be related with plasma fluidity, membrane viscosity, and serum viscosity. Given this extensive literature, measuring viscosity in vivo has important diagnostic and prognostic implications, making it a promising functional imaging application for MPI.

The relaxation behaviour of nanoparticles under zero-field (i.e., when an applied magnetic field is removed) is modeled by two different mechanisms: Neel relaxation where the magnetic moment of the nanoparticle aligns itself internally with the applied field, and Brownian relaxation where the nanoparticle physically rotates to align its magnetic moment with the applied field. Extensive simulations showed the influence of particle parameters such as size and shape anisotropy, magnetization dynamics, and the sequence of applied fields (Weizenecker *et al* 2012, Graeser *et al* 2016). Theoretical modeling for Neel relaxation via Landau-Lifshitz-Gilbert equation and for Brownian relaxation via Fokker-Planck equation were also presented (Rogge *et al* 2013, Reeves and Weaver 2012). Due to the physical rotation of the nanoparticle, Brownian relaxation is directly influenced by the external environment, such as the viscosity of the medium. With this fact in mind, the monitoring of viscosity through the ratio of nanoparticle magnetization harmonics was proposed, with potential extensions to system-matrix based MPI reconstruction (Rauwerdink and Weaver 2010b, Weaver and Kuehlert 2012). The relaxation induced signal delays in x-space based MPI reconstruction were also suggested as a possible tool for estimating the mobility of nanoparticles (Wawrzik *et al* 2013). Recent "color MPI" studies have shown the capability of MPI to separate signals from different nanoparticles (Rahmer *et al* 2015, Hensley *et al* 2015), which could in potential be applied to differentiate nanoparticles in different viscous environments. In these color MPI techniques, a successful separation relied on extensive calibration acquisitions that essentially characterize the behaviour of the nanoparticles under different settings, or measurements at multiple drive field amplitudes to reveal the differences between the responses of nanoparticles. MPI cell tracking experiments have also shown that nanoparticles internalized into the cells displayed reduced resolution (Zheng *et al* 2015) or had altered signal characteristics when compared to nanoparticles in water (Them *et al* 2016), which could stem from viscosity effects on the relaxation behaviour of the nanoparticles. Hence, understanding the effects of viscosity on MPI signal is vital in developing the viscosity mapping capabilities of MPI for functional imaging purposes.

In this work, we propose a viscosity mapping technique for MPI that relies on the relaxation induced changes in the nanoparticle response. We characterize these changes through an estimation of the relaxation time constant, without any a priori information on the nanoparticles. With extensive experiments in an in-house magnetic particle spectrometer (MPS) device, we show the drive field strength and frequency dependence of the relaxation time constant as a function of viscosity. The results provide guidance on MPI imaging parameters that can accentuate the signal differences between nanoparticles in environments with different viscosities. The proposed technique can be extended to imaging applications, facilitating the viscosity mapping potential of MPI.

## 2. Theory

In x-space MPI reconstruction, the ideal signal (also called the adiabatic signal) is denoted as (Goodwill and Conolly 2010);

$$s_{ideal}(t) = B_1 m \rho(x) * \dot{\mathcal{L}}[kGx]\big|_{x=x_s(t)} kG\dot{x}_s(t) \qquad (2.1)$$

where $B_1$ [T/A] is receiver coil sensitivity, $m$ [Am$^2$] is magnetic moment of the nanoparticle, $\rho(x)$ [particles/m$^3$] is density of nanoparticles along the x axis, $k$ [m/A] is a nanoparticle property, $x_s(t)$ is the time-dependent position of the field free point (FFP), and $G$ is the selection field gradient strength [T/m/$\mu_0$]. In practice, the received MPI signal is affected by the relaxation behaviour of the SPIOs. Previous studies have modeled this effect as a temporal convolution of the ideal MPI signal with an exponential relaxation kernel (Croft *et al* 2012, 2016):

$$s(t) = s_{ideal}(t) * r(t) \qquad (2.2)$$

where,

$$r(t) = \frac{1}{\tau} e^{-\frac{t}{\tau}} u(t) \qquad (2.3)$$

Here, $\tau$ is called the relaxation time constant and $u(t)$ is the Heaviside step function. It has been verified via extensive experimental work that this phenomenological model accurately characterizes the MPI response for a wide range of frequencies and drive field amplitudes (Croft *et al* 2012, 2016).

In MPI, positive and negative half cycles of the drive field move the FFP forward and back across the scanned partial field-of-view (FOV), repetitively. As a result of this back and forth scanning process, the ideal MPI signal acquired during the positive and the negative scanning directions are "mirror symmetric", independent of the nanoparticle distribution $\rho(x)$. Here, mirror symmetry refers to the positive half cycle and the time-reversed and negated negative half cycle of the signal being identical. In the case of relaxation, however, the MPI signal is effectively "blurred" along the scanning direction, which in turn breaks the mirror symmetry between the two half cycles (Onuker and Saritas 2015, Muslu *et al* 2016). In theory, the relaxation time constant can be estimated from the half cycles of the MPI signal using the mirror symmetry of the underlying ideal MPI signal, without any a priori information about the nanoparticle type or distribution. The mathematical modeling behind this process is explained in detail in appendix A1. In this work, we apply this model to the nanoparticle signal from an MPS setup, with the goal of measuring the viscosity through estimating the relaxation time constant.

## 3. Materials and Methods

*3.1. Experimental Setup:*

Figure 1 shows the overall experimental configuration. Our in-house magnetic particle spectrometer (MPS) setup (also called an MPI relaxometer) had a drive coil that generates 0.97 mT/A magnetic field with 95% homogeneity in a 7-cm-long region down its bore (Utkur and Saritas 2015). The receive coil was a three-section gradiometer-type coil that allowed decoupling of the transmit and receive coils. This receive coil design had 7 layers of Litz wire, where each layer contained 41 turns in the middle and 21 turns on both sides. The measurement chamber inside the receive coil allowed a sample tube of 1.5 cm in length and 0.8 cm in diameter. According to our previous simulations, due to the symmetry in the design, an MPS setup that incorporates a three-section receive coil can utilize a shorter drive coil with 23% reduced power consumption to achieve results similar to a design that uses a two-section gradiometer-type receive coil (Utkur and Saritas 2015). The self-resonance of the receive coil was measured at around 250 kHz. The drive field amplitude was calibrated via a Hall effect gaussmeter (LakeShore 475 DSP Gaussmeter) and monitored in real time via a Rogowski current probe (LFR 06/6/300, Power Electronic Measurements Ltd).

The drive coil was impedance matched to an AC power amplifier (AE Techron 7224) using a capacitive network, which enabled maximum transfer to the load while low-pass filtering potential higher harmonics of the drive field. On the receive side, the nanoparticle signal was amplified with a low-noise voltage preamplifier (Stanford Research Systems SR560). The MPS setup was controlled with an in-house MATLAB script (Mathworks, Natick, MA). A data acquisition card (National Instruments, NI USB-6363) sent the drive field signal to be amplified to the power amplifier and digitized the signal from the receive chain. During the experiments, ambient temperature inside the measurement chamber was monitored using a fiber optic temperature probe (Neoptix Reflex-4).

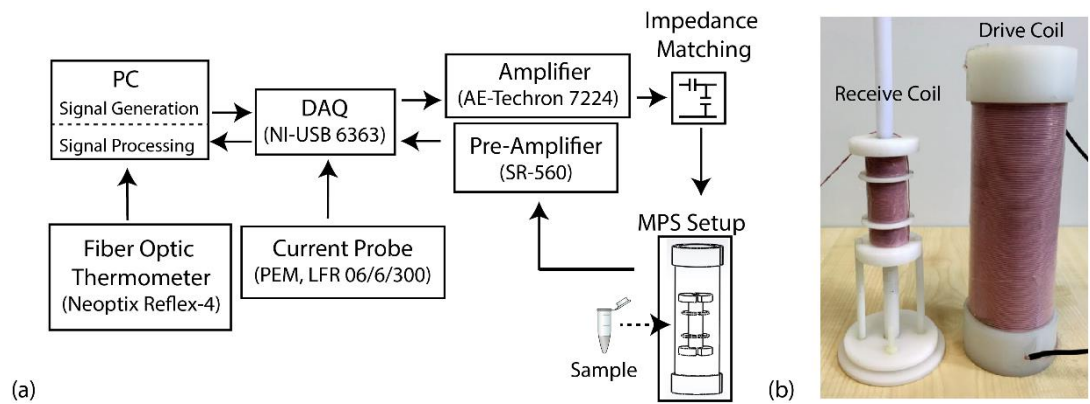

**Figure 1.** Overview of the experimental setup. (a) Arrows in the schematic denote the workflow of the transmit/receive chain for our in-house magnetic particle spectrometer (MPS) setup, which is controlled via a data acquisition card (DAQ) through MATLAB. A fiber optic thermometer and a current probe provide real-time monitoring of the temperature and the magnetic field in the MPS setup, respectively. (b) A picture of the receive coil and the drive coil. The receive coil is placed axially inside the drive coil and its vertical position is adjusted via the pedestal beneath, to minimize the mutual inductance between the two coils.

*3.2. Sample Preparation and Experimental Procedures:*

Viscosity in the cell cytoplasm for an aqueous phase is reported to be around 1-2 mPa•s (close to the 0.89 mPa•s viscosity of pure water at 25°C), and this number can become larger as the activity inside the cell increases (Kuimova *et al* 2009). Viscosity of the blood, on the other hand, is reported to be in the range of 1.3-7.8 mPa•s for hematocrit percentage in the total blood volume ranging between 14-76% (Pirofsky 1953). To provide biologically relevant results in light of this information, we prepared two different sets of samples:

*1.1.1.   Sample Set #1.* 11 samples with viscosities ranging between 0.89 mPa•s and 15.33 mPa•s. The total volume and nanoparticle amount in each sample was kept identical to avoid any Fe concentration bias in the results. Accordingly, each sample contained 50 μL of undiluted nanomag-MIP (Micromod GmbH, Germany) nanoparticles with 89 mmol Fe/L concentration. These samples were then diluted to a total volume of 170 μL, with varying mixtures of deionized water and glycerol to reach the target viscosity levels at 25°C (Sheely 1932). Here, 0.89 mPa•s corresponds to the viscosity of pure water, whereas 15.33 mPa•s corresponds to 68% glycerol by volume.

*1.1.2.   Sample Set #2.* To investigate the biological range in more detail, 20 samples with viscosity levels ranging from 0.89 mPa•s to 3.97 mPa•s were prepared following a procedure similar to the one described above. For this second set, the total sample volume was 200 μL, starting from an initial 80 μL volume of undiluted nanomag-MIP nanoparticles. Here, 3.97 mPa•s corresponds to the viscosity of 45% glycerol by volume.

*1.1.3.   Sample Set #3.* To compare the viscosity effect on different nanoparticles, 11 samples were prepared using VivoTrax ferucarbotran nanoparticles (Magnetic Insight Inc., USA) with 98 mmol Fe/L concentration. Note that this nanoparticle has the same chemical composition as Resovist. For this set, same viscosity levels (i.e., ranging from 0.89 mPa•s to 15.33 mPa•s) and volumes were used as in sample set #1.

Samples in set #1 were tested at four different drive field frequencies; 250 Hz, 550 Hz, 1.1 kHz, and 10.8 kHz, and under three different drive field peak amplitudes; 5 mT, 10 mT, and 15 mT. Samples in set #2 were tested at the same four frequencies, but only at 15 mT field amplitude. Samples in set #3 were tested at 550 Hz and 1.1 kHz, under two different drive field peak amplitudes: 10 mT and 15 mT. For each experiment, the amplified signal from the receive chain was digitized with 2 MS/s bandwidth. Acquisition lengths were chosen such that each cosine excitation pulse contained at least 35 periods. For each measurement, the mean of 16 acquisitions were recorded to increase signal-to-noise ratio (SNR). This step was performed first with no sample inside the measurement chamber to serve as the background measurement, and then with the sample. Overall, each experiment was repeated three times.

*3.3. Data Preprocessing:*

Acquired data sets were processed in MATLAB. For each experiment, the background signal was subtracted from the nanoparticle signal to remove the direct feedthrough and other potential interferences. An initial frequency selection was performed by choosing the higher harmonics of the drive field frequency in Fourier domain and setting all other frequencies to zero. This step removed the remaining direct feedthrough signal at the fundamental frequency, as well as the noise in non-harmonic frequencies that would otherwise reduce signal quality. A subsequent high-order zero-phase digital low-pass filter (LPF) was applied in time domain. The cut-off frequency of LPF was determined via comparing the signal power at harmonic frequencies with the noise power at non-harmonic frequencies, which was computed before setting those frequencies to zero. Accordingly, the cut-off frequencies corresponded to $40^{th}$, $50^{th}$, $50^{th}$, and $18^{th}$ harmonics for 250 Hz, 550 Hz, 1.1 kHz, and 10.8 kHz drive fields, respectively. The usage of fewer harmonics at 10.8 kHz was to avoid the self-resonance frequency of the receive coil, which contaminated signals above 200 kHz.

*3.4. Relaxation Time Constant Estimation:*

The acquired signal contains both the relaxation effects, as well as a phase lag due to inductive/capacitive circuitry. Note that for the ideal MPI signal, the signal peaks would occur at central time points of half cycles, which would trivialize the phase lag estimation. In the case of relaxation, however, the signal peaks are inherently delayed in time with respect to the half-cycle centers.

Here, the relaxation time constant and the system delays were estimated simultaneously from the received signal, without any prior information. Accordingly, a sweep of phase lags, $\phi$, between 0 and $\pi$ were applied to the received signal and the signal was divided into positive/negative half cycles for each phase lag case. The relaxation time constant, $\tau$, was estimated from these half cycles as described in appendix A1. The signal was then Wiener deconvolved by the relaxation kernel $r(t)$ given in (2.3). A necessary condition for the successful implementation of the deconvolution step is to have sufficiently high SNR (Bente *et al* 2015), which was achieved via the data processing steps outlined in section 3.3. Next, the root-mean-squared error (RMSE) between the deconvolved positive half cycle, $\hat{s}_{\text{pos,ideal}}$, and mirrored negative half cycle, $\hat{s}_{\text{neg,ideal,mirr}}$, was computed. This step was to check whether the mirror symmetry was restored after deconvolution. Finally, the $\{\tau, \phi\}$ pair that minimized RMSE was chosen as the solution, i.e.,

$$\{\hat{\tau}, \hat{\phi}\} = \text{argmin}_{\{\tau,\phi\}} \int \left|\hat{s}_{\text{pos,ideal}}(t) - \hat{s}_{\text{neg,ideal,mirr}}(t)\right|^2 dt \qquad (3.1)$$

The restoration of mirror symmetry via this technique was confirmed via visual inspection of results on randomly selected experimental data. The tails of the half cycles were excluded from the RMSE calculation step, as they corresponded to the lowest SNR regions of the signal. Accordingly, 15% of the signal was cut from both sides of the half cycles during RMSE calculation. This choice is further explained in appendix A2.

Finally, as each experiment had three repetitions, the mean value and standard deviations across these repetitions were computed for each viscosity level at each drive field strength and frequency.

## 4. Results

Overall, 636 distinct experiments were performed. For sample set #1, nanoparticles at 11 viscosity levels were measured at 3 different field strengths, at 4 different frequencies, and with 3 repetitions (396 experiments). For sample set #2, nanoparticles at 20 viscosity levels were measured at a single field strength, at 4 different frequencies, and with 3 repetitions (240 experiments).

First, the effects of viscosity on the MPS signal were investigated at different drive field amplitudes and frequencies. Figure 2(a) compares two cases of MPS signals: at viscosity levels of 0.89 mPa•s (nanoparticles in water) and 15.33 mPa•s (nanoparticles in water/glycerol mixture of 68% glycerol by volume), both acquired at 1.1 kHz and 15 mT drive field. Considering that both samples had the same iron content, one can note that the signal at low viscosity level is reduced when compared to that at high viscosity. In addition, the response at low viscosity is visibly wider, indicating an increased relaxation time constant. The estimated time constants were $\tau$ = 33 µs and $\tau$ = 24 µs for $\eta$ = 0.89 mPa•s and $\eta$ = 15.33 mPa•s, respectively. Figure 2(b) compares the same samples at 10.8 kHz and 15 mT drive field. This time, in contrast to the results at 1.1 kHz, the signal at high viscosity is slightly wider than the one at low viscosity. The estimated times constants were $\tau$ = 3.5 µs and $\tau$ = 3.6 µs for $\eta$ = 0.89 mPa•s and $\eta$ = 15.33 mPa•s, respectively.

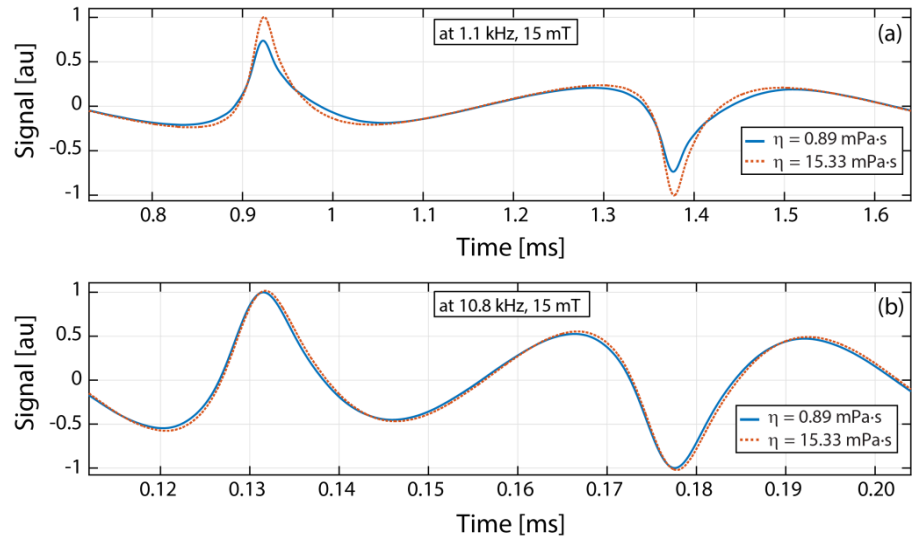

**Figure 2.** Effects of viscosity on nanoparticle signal. (a) The measured signal of nanomag-MIP nanoparticles at 1.1 kHz and 15 mT drive field are shown for low viscosity (0.89 mPa•s) and high viscosity (15.33 mPa•s) cases. The low viscosity case displays a wider response with reduced peak amplitude. The calculated relaxation time constant for these two cases were $\tau = 33$ μs and $\tau = 24$ μs, respectively. (b) The same samples were measured at 10.8 kHz and 15 mT drive field, yielding $\tau = 3.5$ μs and $\tau = 3.6$ μs, respectively. Here, the high viscosity case has a slightly wider response than the low viscosity case.

The relaxation times constants were estimated for all MPS signals acquired. The results for sample set #1 (with 11 different viscosity levels) at four different drive field frequencies and three different drive field strengths are presented in figures 3 and 4. Error bars denote the mean estimated time constants and standard deviations over three repetition experiments. In figure 3, for fixed frequency, estimated time constants decrease as field strength increases. The relaxation time constant as a function of viscosity displays a non-monotonic trend. Especially at low drive field frequencies and amplitudes (e.g., at 250 Hz and 5 mT), $\tau$ first increases sharply and then decreases with increasing viscosity, finally converging to a roughly constant value. At 10.8 kHz, on the other hand, $\tau$ slowly but steadily increases with increasing viscosity.

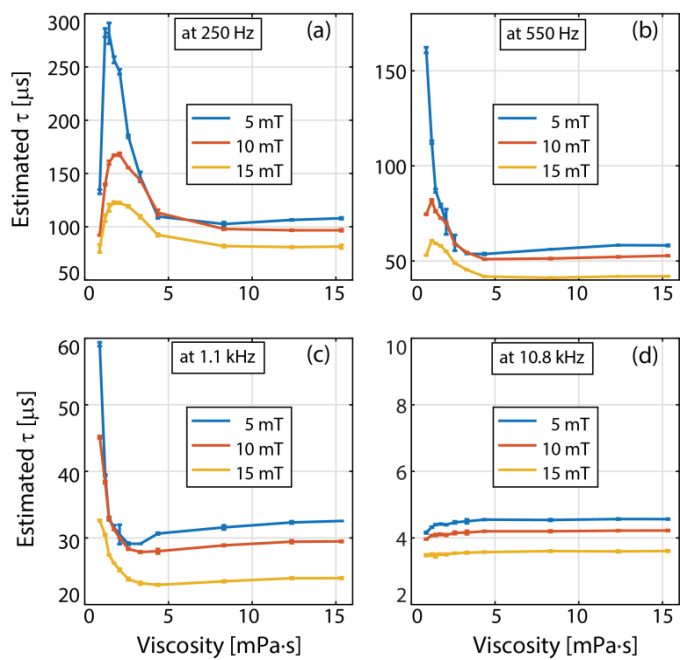

**Figure 3.** Relaxation time constants as a function of viscosity, at four different drive field frequencies for sample set #1 (nanomag-MIP at 11 different viscosities ranging between 0.89 mPa•s to 15.33 mPa•s). Error bars denote the mean values and standard deviations over three repetition experiments.

In figure 4, the same data are re-plotted, this time to highlight the effect of drive field frequency on estimated time constants. In figure 4(a) to 4(c), the time constants decrease with increasing frequency. Interestingly, one can see that there is a global trend in these curves, such that the same τ vs. viscosity curve is scaled down and shifted towards the left as frequency increases. Figure 4(d) to 4(f) presents these results with a normalized time axis, where the estimated time constants are normalized by the period of the drive field at each frequency, and multiplied by 100. Effectively, this unified y-axis gives an indication of the percentage effect of relaxation delays at each frequency. Overall, for fixed drive field amplitude, the relaxation effects are more prominent at higher frequencies. Exceptions to this trend are low viscosity and low drive field strengths, where the trends can be reversed (e.g., at 5 mT for viscosities lower than 3 mPa•s).

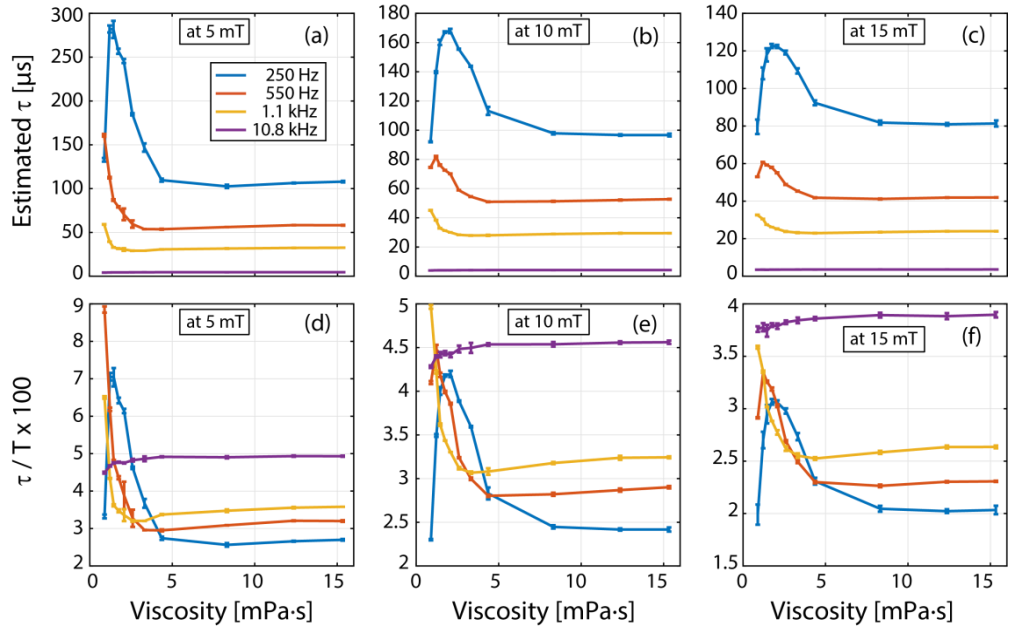

**Figure 4.** Relaxation time constants as a function of viscosity, at three different drive field strengths for sample set #1 (nanomag-MIP at 11 different viscosities ranging between 0.89 mPa•s to 15.33 mPa•s). Error bars denote the mean values and standard deviations over three repetition experiments. (a) to (c) display the same data as in figure 3, re-plotted to emphasize the effect of drive field frequency at a fixed field strength. (d) to (f) display the results in a unified y-axis, where the estimated time constants are normalized by the period of the drive field at each frequency, multiplied by 100.

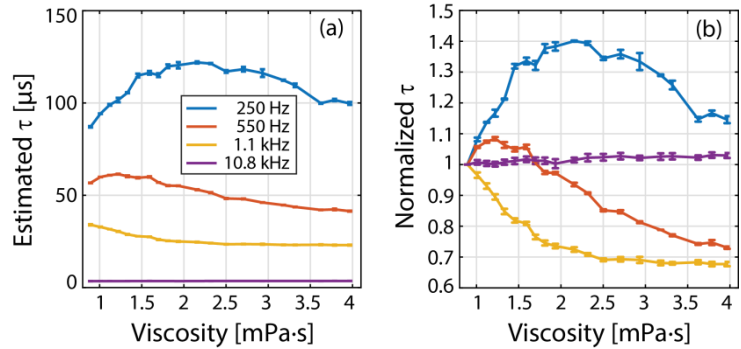

**Figure 5.** Relaxation time constant vs. viscosity for the biologically more relevant range of viscosities, for sample set #2 (nanomag-MIP at 20 different viscosities ranging between 0.89 mPa•s to 3.97 mPa•s), measured at four different frequencies at 15 mT field strength. Error bars denote the mean values and standard deviations over three repetition experiments. (a) Estimated time constants and (b) time constants normalized by that of nanoparticles in water to highlight the percentage change in τ vs. viscosity.

To investigate the biologically more relevant viscosities of up to 4 mPa•s, the time constants for sample set #2 with 20 different viscosity levels between 0.89 mPa•s and 4 mPa•s were estimated at four different drive field frequencies with 15 mT field strength. Another motivation for this second set of experiments was to ensure that the τ vs. viscosity curves did not exhibit erratic jumps with a slight variation in viscosity. The results given in figure 5(a) are consistent with the low viscosity portions of the results in figure 4(c), as no erratic behaviour is observed. Note that the estimated time constants at 2.5 mPa•s and 3.6 mPa•s showed slight deviations from the overall trends. These deviations could stem from pipetting errors during the mixing of water and glycerol, as a small variation in glycerol percentage can significantly alter the viscosity level of the mixture. In figure 5(b), the relaxation time constants are normalized by that of nanoparticles in water (i.e., the lowest viscosity of 0.89 mPa•s), so that the percentage change in τ can be visualized. At 250 Hz, τ first increases by 40% at around 2.2 mPa•s, then gradually decreases. At 1.1 kHz, τ monotonically decreases, showing a 32% reduction at around 4 mPa•s. At 10.8 kHz, on the other hand, τ increases by only 2% within this viscosity range.

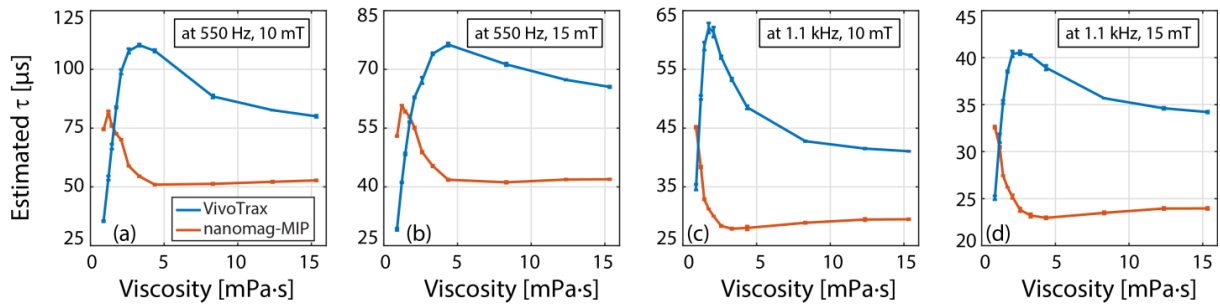

**Figure 6.** Effects of viscosity on the relaxation time constants of two different nanoparticles. VivoTrax and nanomag-MIP nanoparticles were measured at four different drive field conditions, and at 11 different viscosities ranging between 0.89 mPa•s to 15.33 mPa•s. Error bars denote the mean values and standard deviations over three repetition experiments.

To provide a better understanding of the effects of viscosity on different nanoparticles, experiments were performed with VivoTrax nanoparticles at 11 different viscosity levels. The results are given in figure 6, in comparison with nanomag-MIP nanoparticles under the same drive field conditions. Accordingly, the estimated time constants for VivoTrax are larger than those for nanomag-MIP, in general. Importantly, the relaxation time constants for VivoTrax display a non-monotonic trend as a function of viscosity, as was the case for nanomag-MIP. Furthermore, the global trend seen in nanomag-MIP is also observed for VivoTrax: the same τ vs. viscosity curve is scaled down and shifted towards the left with increasing frequency, as can be seen comparing figure 6(a) with 6(c).

During all experiments, the ambient temperature inside the measurement chamber remained between 23-36°C, which corresponds to a minor 4% variation in absolute temperature in Kelvins. This variation was due to resistive heating of the drive coil under currents reaching 22 A, which causes a power dissipation of about 350 Watts. Nonetheless, during signal acquisition, each sample tube stayed in the measurement chamber for only 5 seconds at a time. In a separate control experiment, we also measured the temperature *inside the sample tubes* for the case when the chamber temperature was at 36°C. Accordingly, during the 5-second interval that the sample was exposed to convective heat transfer, its temperature only changed by 0.3-0.4°C. Hence, we do not expect any temperature bias for the results reported in this work.

## 5. Discussion

Viscosity sensing is currently infeasible with ultrasound, magnetic resonance imaging, or other preclinical imaging techniques. This work demonstrated the potential of MPI for viscosity mapping through the estimation of the nanoparticle relaxation time constant. As shown in the results in figures 3 to 5, the trend in the τ vs. viscosity curves strongly depend on the drive field strength and frequency. Ideally, a quantitative viscosity mapping technique should provide a one-to-one mapping of the measured parameter to the viscosity level. Such a technique should also show a sufficiently large variation of the measured parameter as a function of viscosity, such that the viscosity can be determined accurately. For the biologically relevant range investigated in figure 5, these requirements are best satisfied by the measurements at 1.1 kHz, where the relaxation time constant monotonically decreases with increasing viscosity and displays greater than 30% variation. At 10.8 kHz, on the other hand, the time constant almost remains constant at different viscosity levels. These findings imply that the regular MPI operating frequencies of 25 kHz or 150 kHz are not optimal for viscosity mapping purposes. For the very low drive field frequencies around 250 Hz, on the other hand, τ is a non-monotonic function of viscosity, which would create an ambiguity in viscosity mapping. A potential solution at those frequencies could be to perform measurements at two different field strengths to determine the viscosity level.

The well-known Brownian relaxation time constant of nanoparticles is given as $\tau = 3V\eta/k_BT$, where the time constant increases linearly with viscosity, which is clearly not the case for the results shown in this work. It should be emphasized that the Neel and Brownian time constants are valid for the zero field case only, i.e., when an applied external magnetic field is suddenly removed. Hence, they do not model the more complex cases such as a sinusoidal external field. For those cases, simulation results solving Landau-Lifshitz-Gilbert equation or Fokker-Planck equation were presented (Rogge *et al* 2013, Reeves and Weaver 2012). A previous method considered a dynamic magnetization model, where the ratio of $5^{th}$ to $3^{rd}$ harmonics of nanoparticle magnetization was utilized to probe viscosity. Accordingly, a non-monotonic trend was observed as a function of viscosity, both in simulations as well as in experiments with Feridex in various gelatin concentrations (Rauwerdink and Weaver 2010b). Here, the relaxation time constant in (2.2) and (2.3) is a lumped parameter that models the blurring effect in the time-domain MPI signal, without considering the underlying physical mechanisms. Still, the result in figure 2(a) that displays a visibly narrower response at higher viscosity level is counter intuitive, as one would expect a slower nanoparticle response with increasing viscosity. This could potentially be due to the chemical differences between the water solution vs. the high viscosity water/glycerol mixture, which can potentially affect the interaction of the nanoparticles with each other and with the medium. Experiments in solutions with similar viscosities but with different chemical properties could help in understanding this effect.

Previous studies have demonstrated that the MPI signal properties strongly depend on the nanoparticle type (Ferguson *et al* 2015). The experiments here were performed using nanomag-MIP and VivoTrax, where both nanoparticles displayed similar non-monotonic trends. Interestingly, the response of VivoTrax at 1.1 kHz and 10 mT was very similar to the response of nanomag-MIP at 550 Hz and 15 mT, as can be seen in figure 6. These similarities imply that the same physical mechanism is taking place for both nanoparticles, albeit at different strengths. It should be noted that both of these nanoparticles are multi-core nanoparticles with clusters made up of smaller nanoparticles (Eberbeck *et al* 2013). It remains to be shown whether similar trends are valid for single-core nanoparticles with large diameters.

One important factor to take into account during viscosity mapping with the proposed technique is the homogeneity of the drive field. As seen in figure 3, for a fixed frequency, changes in drive field amplitude significantly change the estimated relaxation time constant. Hence, to ensure a quantitative mapping of the viscosity during 3D imaging applications, either the drive field needs to be highly homogeneous within the field-of-view (FOV), or one needs to know the field map within the FOV. Another important consideration is the human safety limits of the applied drive fields (Saritas *et al* 2013b, Schmale *et al* 2013, Saritas *et al* 2015). According to the results shown in this work, frequencies around 1 kHz have the potential to provide a one-to-one viscosity mapping capability. For 1 kHz, the magnetostimulation safety limit in the human torso can roughly be estimated as 20 mT-peak (Saritas *et al* 2013b). Hence, the results presented at 1.1 kHz and 15 mT-peak field strength are actually applicable according to the human safety limits of MPI. It should be noted that operating at a lower frequency than the current MPI frequencies (25 kHz, or lately 150 kHz) would have both advantages and disadvantages in terms of image quality. Previous work has shown that peak signal values changed approximately linearly with the drive field frequency and amplitude (Weber *et al* 2015, Croft *et al* 2016). Similarly, in our experiments, the received signal strength (before the low-noise preamplifier) at 1.1 kHz was around 10 times lower than that at 10.8 kHz for the same drive field amplitude (results not shown). Hence, the image SNR would decrease when operating at lower frequencies. On the other hand, lower drive field frequencies could yield better resolution images, as previously demonstrated via resolution measurements in a relaxometer setup (Croft *et al* 2016). Likewise in our experiments, the nanoparticle signal at 1.1 kHz had a narrower peak than that at 10.8 kHz for the same drive field amplitude, as seen in figure 2.

The exponential model in (2.2) and (2.3) was first proposed and validated for frequencies between 4.4 kHz and 25 kHz (Croft *et al* 2012, 2016). One assumption inherent in this model is that the ideal nanoparticle response without relaxation effects is symmetric with respect to applied field (i.e., the signal peaks occur at central time points of half cycles), which may not be the case for nanoparticles with shape anisotropy or for ferromagnetic nanoparticles. Even in those cases, however, the proposed method could potentially detect the "relative" changes in delay caused by lower/higher viscosity environments. For the multi-core nanoparticles tested here, our analysis shows that the exponential model provides a good fit to the experimental data at 1.1 kHz and 10.8 kHz. At lower drive field frequencies, however, we see that the relaxation deconvolved signal does not fully revert to a mirror symmetric signal. In appendix A2, we analyzed the effect of truncating the tails of the half cycles during $\tau$ estimation. In theory, if the model worked perfectly, it should yield identical $\tau$ independent of truncation percentage. This, however, is not the case at low drive field amplitudes and frequencies. These results indicate that the exponential model provides a better fit to the measured signal at frequencies in the kHz range.

At all frequencies tested in this work, the relaxation time constants showed an asymptotic convergence for viscosity levels above 5 mPa•s, remaining almost flat above 8 mPa•s. These results indicate that viscosity mapping with this technique may not be feasible at very high viscosity levels. However, there is good evidence in the literature that viscosities below 5 mPa•s are biologically more relevant. For example, it has been previously reported that hematocrit level exceeding 46% (>2.8 mPa•s) is considered as an important risk factor for cerebral infarction, atherosclerosis in the arteries, and hypertension (Tohgi *et al* 1978). In another study, elevated viscosity levels from 1.48 mPa•s to 1.71 mPa•s due to myocardial infarction was reported (Fuchs *et al* 1990). These viscosity levels are within the mapping capabilities of the proposed technique, demonstrating the potential of MPI imaging to provide critical functional information.

## 6. Conclusion

In this work, we have demonstrated the viscosity mapping capability of MPI through proof-of-concept experiments in an MPS device. The proposed technique takes advantage of the underlying mirror symmetry in the time-domain MPI signal to blindly estimate a relaxation time constant. The experimental results at various drive field frequencies suggest that a relatively low drive field frequency of around 1 kHz is promising for a one-to-one viscosity mapping. Future imaging applications exploiting the MPI signal's dependence on local viscosity of the nanoparticles will benefit from the results provided in this work.

**Acknowledgements**

The authors would like to thank Akbar Alipour for his assistance in the experimental procedures, and to Tarik Reyhan for valuable discussions. This work was supported by the Scientific and Technological Research Council of Turkey through TUBITAK Grants (114E167), by ERA.Net RUS PLUS Program (TUBITAK 215E198), by the European Commission through FP7 Marie Curie Career Integration Grant (PCIG13-GA-2013-618834), by the Turkish Academy of Sciences through TUBA-GEBIP 2015 program, and by the BAGEP Award of the Science Academy.

**Appendix A.**

**A.1. Relaxation time constant estimation**

A sinusoidal drive field superimposed to a piecewise constant focus field results in a so called "linear" MPI trajectory, which scans a partial field-of-view (FOV) back and forth, repeatedly. With this trajectory, the ideal MPI signal (i.e., the signal without relaxation effects) demonstrates what can be called a "mirror symmetry" between signal parts acquired during the positive and the negative scanning directions. Accordingly, these ideal half-cycle signals can be expressed as follows:

$$s_{pos,ideal}(t) = -s_{neg,ideal}(-t) = s_{half}(t) \tag{A.1}$$

Here, "mirroring" is defined as time reversal and negation of a signal. Previously, MPI signal with relaxation has been modeled as a convolution of nanoparticles' Langevin response with a relaxation kernel, as given in (2.2) and (2.3). In principle, the relaxation process blurs the signal along the scanning direction, breaking the mirror symmetry between the two half cycles. To demonstrate this point, figure A1(a) and A1(b) display the lost mirror symmetry for a measured MPS signal acquired at 1.1 kHz and 15 mT drive field for nanomag-MIP nanoparticles. In theory, the half cycles shown in figure A1(b) would overlap exactly if the nanoparticles did not exhibit any relaxation behaviour.

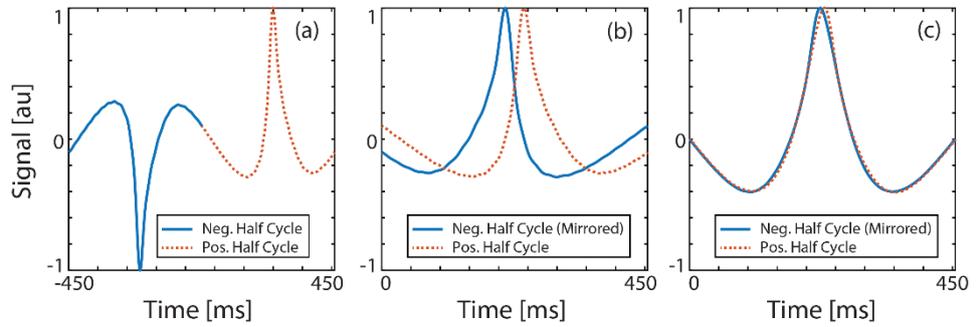

**Figure A1.** Estimating the relaxation time constant directly from the measured MPS signal. (a) Measured MPS signal for nanomag-MIP nanoparticles at 1.1 kHz and 15 mT drive field, for the case of a low viscosity solution (approximately 1.22 mPa•s). The relaxation effect on the signal is clearly visible. (b) The positive half cycle and the mirrored (i.e., time reversed and negated) negative half cycle are overlaid. If the signal had not contained any relaxation effects, these two half cycles would overlap exactly. Due to relaxation, however, each half cycle is "blurred" along the scanning direction. The time constant estimation scheme described in appendix A1 yields τ = 32.9 μs. (c) The measured signal can then be deconvolved by the relaxation kernel with the estimated time constant, which reveals the underlying mirror symmetry.

Using the convolution-based formulation for the MPI signal, the half cycles in the case of relaxation can be expressed as:

$$s_{pos}(t) = s_{pos,ideal}(t) * r(t) = s_{half}(t) * r(t) \tag{A.2}$$

$$s_{neg}(t) = s_{neg,ideal}(t) * r(t) = -s_{half}(-t) * r(t) \tag{A.3}$$

Note that $s_{half}(t)$ is the ideal half-cycle signal. Using the exponential model for $r(t)$, these two equations can be solved simultaneously to reveal both τ and $s_{half}(t)$ (Onuker and Saritas 2015, Muslu

*et al* 2016). Here, we are particularly interested in the relaxation time constant τ, which can be expressed in closed form in the Fourier domain:

$$\mathcal{F}\{r(t)\} = R(f) = \frac{1}{(1+i2\pi f\tau)} \tag{A.4}$$

$$\mathcal{F}\{s_{\text{pos}}(t)\} = S_{\text{pos}}(f) = S_{\text{half}}(f) \cdot R(f) \tag{A.5}$$

$$\mathcal{F}\{s_{\text{neg}}(t)\} = S_{\text{neg}}(f) = -S_{\text{half}}(-f) \cdot R(f) \tag{A.6}$$

Here, (A.4) is the known 1D Fourier transform of an exponential function and (A.6) follows from time reversal property of Fourier transform. Since $s_{\text{half}}(t)$ is real valued, its Fourier transform $S_{\text{half}}(f)$ has conjugate symmetry property. Accordingly, the last equation can be re-written as

$$\mathcal{F}\{s_{\text{neg}}(t)\} = S_{\text{neg}}(f) = -S_{\text{half}}^*(f) \cdot R(f) \tag{A.7}$$

where the superscript star sign denotes the conjugation operation. Finally, combining (A.4) to (A.7), we can express the relaxation time constant as follows:

$$\tau = \frac{S_{\text{pos}}^*(f) + S_{\text{neg}}(f)}{i2\pi f(S_{\text{pos}}^*(f) - S_{\text{neg}}(f))} \tag{A.8}$$

Once this time constant is estimated, the signal can be deconvolved by the relaxation kernel, *r(t)*, to reveal the underlying ideal MPI signal. In theory, this deconvolution steps recovers the mirror symmetry between the half cycles. This process is demonstrated in figure A1(c).

In this work, the relaxation time constant and system delays were estimated simultaneously (see section 3.4). The {τ, ϕ} pair that minimized the RMSE between the deconvolved versions of the positive and the mirrored negative half cycles was chosen as the solution.

### A.2. Influence of tail truncation on estimated time constants

The model in (2.2) and (2.3) is phenomenological, i.e., it is based on extensive experimental work that demonstrates a good fit to this model (Croft *et al* 2012). If this model could mimic the relaxation behaviour perfectly, deconvolution with the correct relaxation kernel would restore mirror symmetry completely. As shown in figure A1(c), the half cycles after deconvolution display close to perfect mirror symmetry, notwithstanding a small region of mismatch at the signal peak locations. The mismatch in the peaks could be avoided by assigning more weight to the goodness-of-fit at the center of half cycles. Doing so, however, would worsen the overlap in the remaining portions of the half cycles.

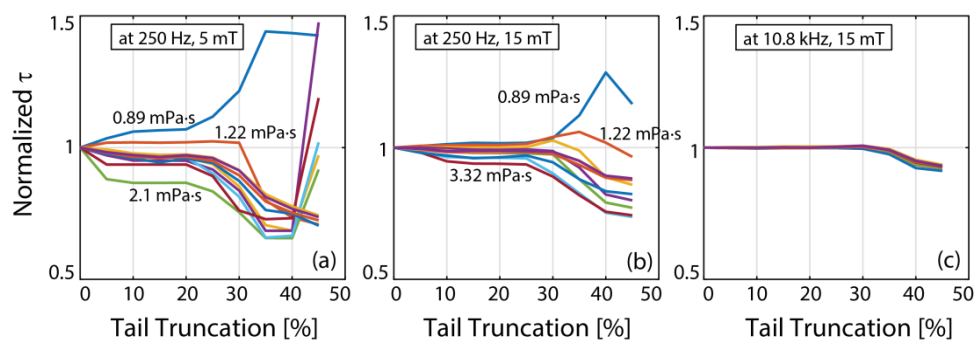

**Figure A2.** Influence of tail truncation on the estimated relaxation time constants. If the relaxation model worked perfectly, these curves would be horizontally flat, yielding constant τ estimation independent of truncation percentage. At low drive field amplitude and frequency as in (a), large truncation percentage yields diverging results, whereas zero truncation shows a slight variation from the expected flat response. In contrast, at high drive field amplitude and frequency as in (c), the estimations remain constant up to 35% tail truncation. In all cases, tail truncations between 10%-20% yielded constant τ estimations. Results are for nanomag-MIP nanoparticles.

We have analyzed the influence of truncating different percentages of the deconvolved half-cycle tails during RMSE calculations. For example, when 100% of the half cycles are utilized (i.e., 0% truncation), the RMSE calculation takes into account the overlap of both the tails and the center of the half cycle. Alternatively, one can truncate a percentage of the tails before calculating the RMSE, such that the overlap of the central regions becomes the main target. Figure A2 shows the effect of tail truncation percentage on the estimated relaxation time constants. The values are normalized to show the overall trends. Here, a 10% truncation indicates 10% of half cycles truncated from both tails, with the remaining central 80% utilized in RMSE calculation. If the relaxation model worked perfectly, one would expect to see a flat τ vs. truncation percent curve, i.e., estimated τ would be independent of the truncation percentage. This is in fact the case at high frequency and high field amplitude, as seen in figure A2(c). At low frequencies and low field amplitudes, however, a large truncation percentage yields diverging results. These results indicate that the exponential model is a better fit at high drive field frequencies and amplitudes. In all cases considered in this work, τ estimation remained flat for truncation percentages ranging between 10% and 20%. Accordingly, a truncation percentage of 15% was utilized in all τ estimations presented in this work.